\documentclass{ws-procs975x65}
\usepackage{graphicx}
\usepackage{ogonek}

\begin{document}

\title{Gravitational lensing by compact objects within plasma}
\author{Adam Rogers$^*$}
\address{Physics \& Astronomy, University of Manitoba,\\
Winnipeg, MB, Canada\\
$^*$E-mail: rogers@physics.umanitoba.ca}

\begin{abstract}
Frequency-dependent gravitational lens effects are found for trajectories of electromagnetic rays passing through a distribution of plasma near a massive object. Ray propagation through plasma adds extra terms to the equations of motion that depend on the plasma refractive index. For low-frequency rays these refractive effects can dominate, turning the gravitational lens into a mirror. While light rays behave like particles with an effective mass given by the plasma frequency in a medium with constant density, an inhomogeneous plasma introduces more complicated behavior even for the spherically symmetric case. As a physical example, the pulse profile of a compact object sheathed in a dense plasma is examined, which introduces dramatic frequency-dependent shifts from the behavior in vacuum.
\end{abstract}

\keywords{Gravitational lensing; Plasma; Neutron stars; Compact objects.}

\bodymatter
\section{Introduction}

The effects of an optical medium in the curved background of general relativity \cite{synge} were first used to describe the deflection of radio waves from distant sources observed through plasma in the solar corona \cite{solarPlasma1, solarPlasma}. Gravitational lensing in the presence of plasma has since become an active and interesting field of investigation \citep{review}, ranging in scope from the weak field \cite{mao14, plasmaLensingKerr} to the strong field regimes \citep{BKT09, BKT10, perlickGR, BB03}.

The effects of plasma on the formation of gravitationally lensed images by black holes or other compact objects are generally difficult to detect due to the small angular size of the images, and require multi-waveband monitoring of microlensing events \citep{plasmaLensingKerr}. As an alternative, we investigate the pulse profiles generated by a compact object embedded in a continuous and smoothly varying refractive medium. In fact, even relatively simple cases of joint gravitational and plasma lensing phenomena can produce dramatic results that differ from the vacuum behavior \citep{rogers15, perlick15, shadow}. Our motivation in this work is to provide a simple example that incorporates novel frequency-dependent effects into general relativity.

\section{Theory}
\label{sec:theory}
We follow the method and notation of Synge (Ref.~\refcite{synge}) to find the trajectories of photons in the presence of an isotropic, inhomogeneous plasma distribution with $G=c=1$. We keep the $\hbar$ factors in the definitions but perform the calculations with $\hbar=1$.

Let us place a slowly rotating compact object of mass $M$ and radius $R$ at the origin. For $r>R$, the space-time is described by the Schwarzschild metric
\begin{equation}
\text{d}s^2=-A(r)\text{d}t^2+A(r)^{-1}\text{d}r^2+r^2\text{d}\theta^2+r^2\sin^2(\theta)\text{d}\phi^2
\end{equation}
with $A(r)= 1 - r_{\text{g}}/r$, $r_{\text{g}}=2M$ and coordinate indices $(t,r,\theta,\phi)$. We use the index $\alpha$ to sum over the spatial coordinates and $i,j$ to run over all four. For simplicity, suppose a distribution of cold, non-magnetized plasma surrounds the mass with refractive index
\begin{equation}
n^2=1- \frac{\omega_{\text{e}}^2}{\omega^2},
\label{nFreq}
\end{equation}
where the frequency along a ray is
\begin{equation}
\omega(r)=\frac{\omega_\infty}{A(r)^\frac{1}{2}}
\end{equation}
in terms of the frequency at infinity $\omega_\infty$. The plasma frequency is
\begin{equation}
\omega_{\text{e}}(r)=\sqrt{ \frac{4 \pi e^2 N(r)}{m} }=\sqrt{ \frac{k}{r^h} }
\label{plasmaFreqDef}
\end{equation}
with $e$ and $m$ the electron charge and mass respectively. The plasma number density is treated as a radial power law $N(r)=N_0/r^h$, with $h \geq 0$. As a physically motivated choice for the examples we set $N_0$ equal to the Goldreich-Julian density \cite{JG69}, and label the constant coefficient as $k$ in Eq.~(\ref{plasmaFreqDef}).

Let us further assume the plasma is static, with four-velocity components $V^{t}=\sqrt{-g^{tt}}$, $V^{\alpha}=0$. While this velocity profile provides a convenient example to illustrate the plasma influence on electromagnetic rays, we note that it is not a realistic accretion flow and will discuss the implications of this choice in Section \ref{sec:effPot}. Using Ref.~\refcite{synge} we relate the components of the photon four-momentum and the refractive index, giving the Hamiltonian in the geometric optics limit:
\begin{equation}
H(x^i,p_i)=\frac{1}{2}\left( g^{ij} p_i p_j + \hbar^2 \omega_{\text{e}}^2 \right) = 0.
\label{nullHamiltonian}
\end{equation}
The position and momenta of photons through the space--time under the influence of gravity and the optical medium are given in terms of the affine parameter $\lambda$ along the trajectory by the partial derivatives
\begin{equation}
\frac{\text{d} x^i}{\text{d} \lambda} = \frac{\partial H}{\partial p_i}
\end{equation}
and
\begin{equation}
\frac{\text{d} p_i}{\text{d} \lambda} = - \frac{\partial H}{\partial x^i}.
\end{equation}
The $r$ and $\phi$ components of the equations of motion can be combined to eliminate $\lambda$, giving an expression for $\text{d} \phi / \text{d}r$. Using the substitution $u=M/r$ and integrating this expression from the surface of the star $r=R$ to a distant observer ($r \rightarrow \infty$), we obtain the maximum visible angle on the stellar surface,
\begin{equation}
\Delta = \int_0^{u_{R}}\frac{\text{d}u}{\left[ \frac{n^2(u)}{x^2} - (1-2u)u^2 \right]^{1/2}},
\label{Delta}
\end{equation}
with $x=b/M$, where $b$ is the impact parameter between a ray and the origin. We describe the impact parameters of the rays which connect the surface of the star with the observer in terms of $\delta$, the angle with respect to the surface normal that a ray is emitted,
\begin{equation}
b=\frac{R n(R)}{ A(R)^{1/2} } \sin \delta
\label{b}
\end{equation}
where the maximum impact parameter $b_{\text{max}}$ occurs for $\delta=\pi/2$. This formula reduces to the standard expression in vacuum when $n=1$ (Refs.~\refcite{pfc}, \refcite{bel02}, \refcite{v1}, \refcite{ve1}). In flat space exactly half the surface of the star is visible ($\Delta_{\text{max}}= \pi /2$). However, gravitational lensing acts to redirect light rays such that a distant observer sees a larger apparent area and more of the stellar surface than would be possible in flat space. In fact, in extreme cases the observer can see portions of the surface multiple times ($\Delta_{\text{max}} > \pi$). The dependence of the impact parameter on the refractive index predicts that the image of the stellar surface shrinks as $n \rightarrow 0$.

\begin{figure}
\centering
\includegraphics[scale=0.85, bb= 83 230 505 540, clip=true]{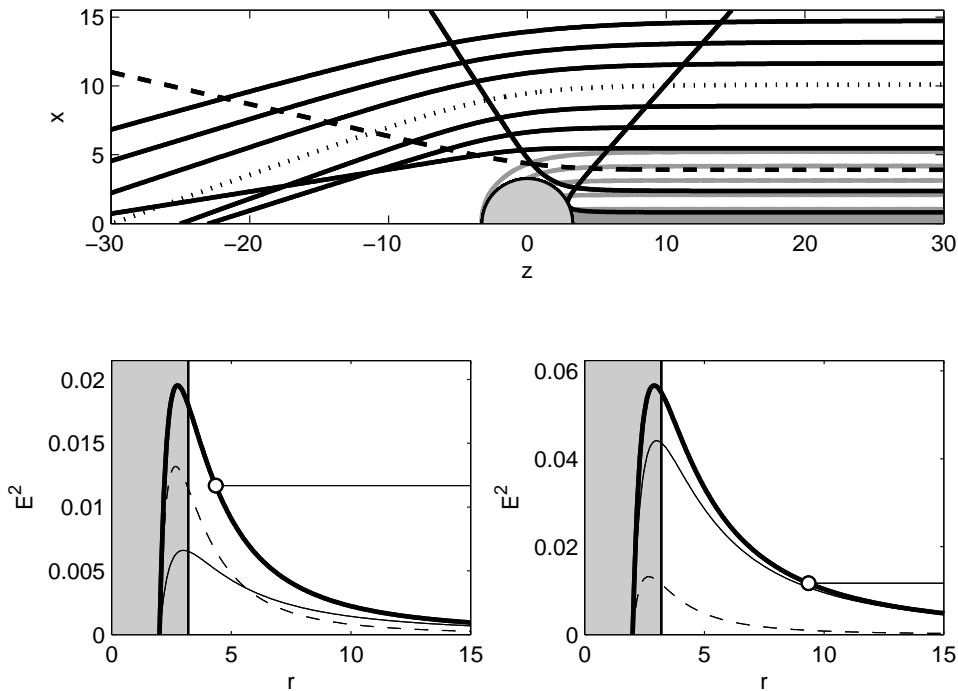}
\caption{Ray--tracing with $\omega^2_{\text{e}}=k/r^3$ for rays near the plasma frequency $\omega_{\text{e}}(R) / \omega(R) = 0.99$, $R/r_{\text{g}}=1.60$. The compact object is shown in the top panel as a light gray half-disk and the free rays ($b>b_{\text{max}}$) are shown in black. Note that none of these trajectories actually touches the surface of the object, and that the closest approach in this figure is slightly exaggerated by the thickness of the line. Rays that intersect the surface $r=R$ ($b<b_{\text{max}}$) are shown as the gray region in the lower right corner. The individual gray curves show the behavior of rays near the object in the vacuum case. The effective potential for the dashed and dotted rays is shown in the lower left and lower right panels respectively. The vacuum case is plotted as a solid curve, and the contribution from plasma is the dashed curve. The total potential is shown as a heavy black curve. The turning point for a given ray is marked by a white disk and the energy is marked by a horizontal line. The interior of the compact object is the gray region on the left-hand side of the panels.}
\label{fig:rayTrace}
\end{figure}

Ray propagation from $R$ to an observer at infinity requires $\omega(R) > \omega_\text{e}(R)$, and absorption occurs if the plasma frequency exceeds the ray frequency. In Fig.~\ref{fig:rayTrace} we show the behaviour of rays with a frequency ratio at the surface of the compact object $\omega_{\text{e}}(R) / \omega(R)=0.99$ and power law index $h=3$. The interior of the star is shown as a gray disk, and free ray trajectories ($b>b_{\text{max}}$) are plotted in black. The solid gray area in the lower right represents rays that intersect the surface of the object ($b<b_{\text{max}}$), and the gray curves show the vacuum case for individual rays near the compact object.

The refractive index $n$ (Eq.~\ref{nFreq}) is small near $R$ due to the dependence of the plasma frequency on density, whereas $r \rightarrow \infty$ gives $n \rightarrow 1$. Since deflection occurs in the direction of increasing $n$, inhomogeneous plasma effects oppose the gravitational deflection. This tension between gravitational (converging) and plasma (diverging) lensing presents interesting dynamics, such that a given type of lensing can dominate depending on the details of the plasma density. Thus even under the influence of gravity, dispersion and refraction within an inhomogeneous plasma may be sufficient to turn a ray counter to the vacuum behaviour.

\section{Effective Potential}
\label{sec:effPot}

The effective potential of the Schwarzschild metric with a power law plasma density is
\begin{equation}
V_{\text{eff}} = \left( 1-\frac{2M}{r} \right)\left( \frac{L^2}{r^2} + \frac{k}{r^h} \right)
\label{effPot}
\end{equation}
for null geodesics. We show two examples of the effective potential plotted against the radial distance from the center of the compact object in the lower panels of Fig.~\ref{fig:rayTrace}. These examples show the potential for the dashed and dotted rays in the top panel of the figure, and explicitly show the contribution due to the optical properties of the plasma. We focus on the $h=3$ case in Fig.~\ref{fig:rayTrace}, however the plasma effects are more pronounced for small $h$.

In general, Eq.~(\ref{effPot}) is analogous to other well-studied cases. For example, $h=0$ gives a constant plasma frequency which acts like an effective mass for photons \cite{kulsrudLoeb, BB03, TBK13}. The $h=2$ case has the plasma term act like an additional contribution to the angular momentum, analogous to the vacuum case with $L^2 \rightarrow L^2 + k$. When $h=3$, $V_{\text{eff}}$ is analogous to the Regge-Wheeler potential \cite{reggeWheeler57}, found from the separable radial part of the Klein-Gordon equation that describes the evolution of scalar-field perturbations on the Schwarzschild background \citep{chandrasekharDetweiller1975}:
\begin{equation}
V_{\text{RW}}=\left( 1-\frac{2M}{r} \right) \left[ \frac{l(l+1)}{r^2}+\frac{2M(1-s^2)}{r^3}\right],
\end{equation}
with the orbital quantum number $l$ and spin given by $s$, along with the condition $l \geq s$ (Ref.~\refcite{LunFackerell74}). The correspondence with the plasma case is exact for spinless particles ($s=0$) with $k=2M$. The $s=1$ case corresponds to electromagnetic perturbations and recovers the vacuum behaviour ($k=0$). For gravitational perturbations $s=2$ and requires $k<0$. This is an unphysical plasma distribution, and describes a medium that has $n>1$ everywhere. These correspondences allow us to recover the circular orbit radii of massless scalar fields with arbitrary spin in the geometric optics limit. Thus, the interaction of passing electromagnetic radiation with a spherically symmetric gravitational field and a plasma density that drops off as $1/r^3$ is analogous to the more general interaction of a massless boson and scalar field perturbation on the Schwarzschild background. This analogy occurs because of the plasma velocity distribution assumed in Section \ref{sec:theory}. Since the plasma is taken to be static it is fixed above the compact object. A different choice of plasma velocity would result in a more complicated Hamiltonian and effective potential.

\section{Effects on Pulse Profiles}
Let us suppose the compact object emits radiation from bright polar cap regions, and that radiation is produced near the surface. For now let us consider a single cap, with center $\theta_0(t)$, with constant angular size $\theta_c$. The bright cap is inclined with respect to the rotation axis by an angle $\xi$ and the rotation axis is inclined to the line of sight by an angle $\chi$. The orientation of the cap is then
\begin{equation}
\theta_0 (t) = \cos ^{-1} \left[ \cos\chi \cos \xi - \sin \chi \sin \xi \cos \gamma_P (t) \right]
\label{theta0Time}
\end{equation}
with the phase $\gamma_\text{P}=\Omega t + \gamma_\text{0}$, rotation frequency $\Omega$ and $\gamma_\text{0}$ an arbitrary phase constant. We assume the rotation frequency is low and the space-time external to the compact object can be reasonably approximated by the Schwarzschild solution.

Relating the observed intensity far from the object $I_{\text{obs}}$ and the emitted intensity from the surface $I_{\text{em}} = n^2(R) I_0=\text{constant}$ through the generalized invariant relationship between specific intensity and frequency along a ray \cite{bicakHadrava}, we obtain
\begin{equation}
I_{\text{obs}}=\left( 1- \frac{2M}{R} \right)^{3/2} I_0
\label{Iobs}
\end{equation}
where we assume the source radiates isotropically. The total flux from the polar cap is given by the product of the intensity and the solid angle of the cap projected onto the observer's sky.

To describe the distortion of the surface features seen by a distant observer, consider a spherical coordinate system fixed with respect to the surface of the star, and a second coordinate system fixed with respect to the observer such that the $\theta=0$ direction points along the line of sight. The solid angle of the polar cap is constant with respect to the coordinates fixed on the stellar surface, but the projection of this solid angle visible on the observer's sky changes as the star rotates. The projection is found using $\Delta$ (Eq.~\ref{Delta}), which takes into account the distortion due to the gravitational lensing and plasma effects. The spherical symmetry of the star allows the solid angle of the polar cap at a given time to be stated in terms of a one-dimensional integral over the impact parameter. We define the width of the cap at a given $\theta$ in terms of a function $H(\theta[x]; \theta_\text{0}, \theta_\text{c} )$, described in detail in Refs. \refcite{pfc} and \refcite{dabrowski}. Taken together, the intensity and solid angle projection give the flux at frequency $\omega$:
\begin{eqnarray}
F_{\omega} = \left(1 -\frac{2M}{R}\right)^{3/2} \frac{M^2}{D^2}\int_0^{ x_\text{max} } I_0  H(\theta[x]; \theta_0, \theta_\text{c} ) x \text{d}x.
\label{Flux}
\end{eqnarray}
The key difference between this result and Ref.~\refcite{pfc} comes from $\Delta$, used to calculate $H( \theta[x]; \theta_\text{0}, \theta_\text{c} )$, and the range of $x$ since the maximum impact parameter now depends on frequency. The presence of an antipodal polar cap simply requires a second component with $\theta_0 \rightarrow \pi-\theta_0$. To generate light curves we recall the time dependence of $\theta_0$ and plot $F_{\omega}$ as a function of the phase $\Omega t$. The specific value of $k$ that appears in the plasma frequency affects only the scale at which plasma effects become significant, and does not change the overall morphology of the calculated pulse profiles.

\begin{figure}
\centering
\includegraphics[scale=0.9, bb= 116 308 480 486, clip=true]{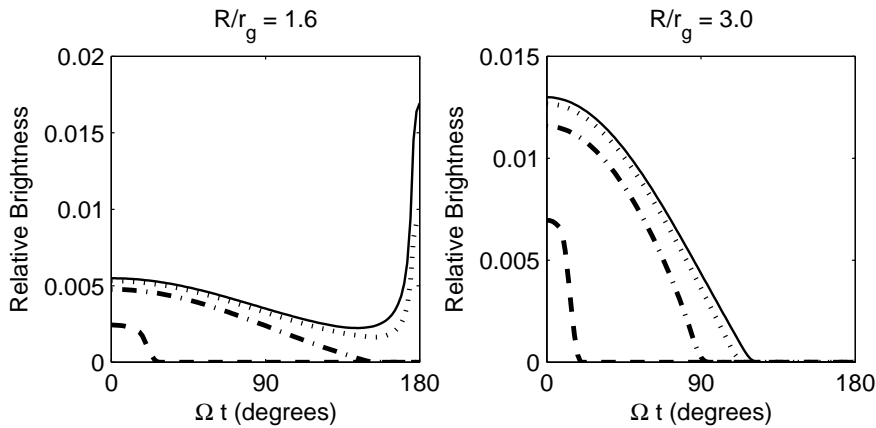}
\caption{Pulse profiles of a single polar cap, $h=3$. The pulses are plotted for an orthogonal configuration with $\chi=\xi=\pi/2$, and frequency ratios $\omega_{\text{e}}(R) / \omega(R)=0.99$ (dashed line), $\omega_{\text{e}}(R) / \omega(R) = 0.60$ (dashed-dotted), $\omega_{\text{e}}(R) / \omega(R) = 0.30$ (dotted) and the vacuum case is the solid line. The aperture angle is $\theta_{\text{c}}=5^{\circ}$. The vacuum case is plotted as a thin, solid line. The left panel has $R/r_{\text{g}}=1.60$ and the right has $R/r_{\text{g}}=3.0$. The pulse profiles narrow as $\omega(R)$ approaches $\omega_{\text{e}}(R)$.}
\label{pulseSingle}
\end{figure}

As an illustrative example, we generate the pulse profile for a single polar cap with a fixed aperture angle $\theta_{\text{c}}=5^{\circ}$ in an orthogonal configuration with $\chi=\xi=\pi /2$. The light curve of a highly relativistic object with $R/r_\text{g}=1.6$ is shown in the left-hand panel of Fig.~\ref{pulseSingle}. The solid line is for $\omega(R) \gg \omega_\text{e}(R)$ and reproduces the vacuum case. This high mass object shows multiple imaging of the polar cap and produces a peak at $\Omega t =0$ and a second, relativistic peak at $\Omega t= \pi$. As the observing frequency is decreased, the relativistic peak becomes less prominent since a smaller part of the surface is visible due to the frequency dependence of the impact parameter in Eq.~(\ref{b}). At low frequencies no portion of the surface is multiply imaged and the profile shows only one peak at $\Omega t=0$. A less extreme example with $R/r_\text{g}=3.0$ is shown in the right panel, and does not display multiple imaging of the surface at any frequency.

\section{Conclusions}

We used the Schwarzschild metric to study the frequency-dependent effects of a cloud of cold, non-magnetized and inhomogeneous plasma surrounding a slowly-rotating compact object. The competition between the converging nature of gravitational lensing and the diverging lens behaviour of the plasma produces interesting orbits for electromagnetic rays and achromatic pulse profiles for compact objects. Despite the simplicity of the assumptions used, the toy model developed here illustrates dramatic shifts from the behaviour in vacuum. We conclude that gravitational lensing together with plasma effects generate a rich set of dynamics for ray propagation in the presence of compact objects.

\section*{Acknowledgments}
I acknowledge and thank Samar Safi-Harb for support through the Natural Sciences and Engineering Research Council of Canada (NSERC) Canada Research Chairs Program. I also acknowledge and thank Oleg Tsupko for inviting me to MG14.


\begin{thebibliography}{0}

\bibitem{synge} J. L. Synge, {\em Relativity: The General Theory}, (North-Holland, Amsterdam, 1960)
\bibitem{solarPlasma1} D. O. Muhleman and I. D. Johnston, {\em Phys. Rev. Lett.}, {\bf 17}, 455 (1966)
\bibitem{solarPlasma} D. O. Muhleman, R. D. Ekers and E. B. Fomalont, {\em Phys. Rev. Lett.} {\bf 24}, 1377 (1977)
\bibitem{review} G. S. Bisnovatyi-Kogan and O. Yu. Tsupko, {\em Plas. Phys. Rep.}, {\bf 41}, 7, (2015)
\bibitem{mao14} X. Er and S. Mao, {\em MNRAS}, {\bf 437}, 2180 (2014)
\bibitem{plasmaLensingKerr} V. S. Morozova, B. J. Ahmedov and A. A. Tursunov, {\em Ap\&SS}, {\bf 346}, 513 (2013)
\bibitem{BKT09} G. S. Bisnovatyi-Kogan and O. Yu. Tsupko, {\em Gravit. Cosmol.}, {\bf 15}, 1 (2009)
\bibitem{BKT10} G. S. Bisnovatyi-Kogan and O. Yu. Tsupko, {\em MNRAS}, {\bf 404}, 1790, (2010)
\bibitem{perlickGR} V. Perlick, {\it Ray Optics, Fermat's Principle and Applications to General Relativity}, (Springer-Verlag, Heidelberg, 2000).
\bibitem{shadow} F. Atamurotov, B. Ahmedov and A. Abdujabbarov, {\em Phys. Rev. D}, {\bf 92}, 8 (2015)
\bibitem{BB03} A. Broderick and R. Blandford, {\em MNRAS}, {\bf 342}, 4 (2003)
\bibitem{rogers15} A. Rogers, {\em MNRAS}, {\bf 451}, 1 (2015)
\bibitem{perlick15} V. Perlick, O. Yu. Tsupko and G. S. Bisnovatyi-Kogan, {\tt arXiv:1507.04217}, {\it preprint}, 2015
\bibitem{JG69} P. Goldreich, W. H. Julian, {\em ApJ}, {\bf 157}, 869 (1969)
\bibitem{pfc} K. R. Pechenick, C. Ftaclas and J. M. Cohen, {\em ApJ}, {\bf 274}, 846 (1983)
\bibitem{bel02} A. M. Beloborodov, {\em ApJL}, {\bf 566}, 2 (2002)
\bibitem{v1} K. S. Virbhadra, {\em Phys. Rev. D}, {\bf 79}, 083004 (2009)
\bibitem{ve1} K. S. Virbhadra and G. F. R. Ellis, {\em Phys. Rev. D}, {\bf 62}, 084003 (2000)
\bibitem{TBK13} O. Yu. Tsupko, G. S. Bisnovatyi-Kogan, {\em Phys. Rev. D}, {\bf 87}, 124009 (2013)
\bibitem{kulsrudLoeb} R. Kulsrud, A. Loeb, {\em Phys. Rev. D}, {\bf 45}, 2 (1992)
\bibitem{reggeWheeler57} T. Regge, J. A. Wheeler, {\em Phys. Rev. D}, {\bf 108}, 4 (1957)
\bibitem{chandrasekharDetweiller1975} S. Chandrasekhar, S. L. Detweiler, {\em Proc. R. Soc. A}, {\bf 344}, 441 (1975)
\bibitem{LunFackerell74} A. W.-C. Lun and E. D. Fackerell, {\em Lett. Nuovo Cimento}, {\bf 9}, 15 (1974)
\bibitem{bicakHadrava} J. Bi\u{c}\'{a}k, P. Hadrava, {\em A\&A}, {\bf 44}, 389 (1975)
\bibitem{dabrowski} M. P. D\k{a}browski and J. Osarczuk, {\em Ap\&SS}, {\bf 229}, 139 (1995)

\end{thebibliography}
\end{document}